\documentclass[a4paper,aps,nofootinbib,onecolumn,notitlepage]{revtex4-1}
\usepackage{amsmath,amssymb}
\usepackage{graphicx}
\usepackage{multirow} 
\begin{document}
\title{Generalised matter couplings in general relativity}

\author{Christian G. \surname{B\"{o}hmer}}
\affiliation{Department of Mathematics and Institute of Origins, University College London, Gower Street, London WC1E 6BT, United Kingdom}
\email{c.boehmer@ucl.ac.uk}

\author{Sante Carloni}
\affiliation{Centro Multidisciplinar de Astrof\'{\i}sica - CENTRA, Instituto Superior Tecnico - IST, Universidade de Lisboa - UL, Avenida Rovisco Pais 1, 1049-001, Portugal}
\email{sante.carloni@tecnico.ulisboa.pt}

\date{\today}

\begin{abstract}
  A new class of modified theory of gravity is introduced where the volume form becomes dynamical. This approach is motivated by unimodular gravity and can also be related to Brans-Dicke theory. On the level of the action, the only change made will be through the volume element which is used in the integration. This is achieved by the introduction of a fourth order tensor which connects the spacetime metric to the new volume form. Using dynamical systems techniques, this model is studied in the context of cosmology. The most interesting result is that there exist parameter ranges where this model starts undergoing an epoch of accelerated expansion, followed by a decelerating expansion which evolves to a final epoch of accelerated expansion. 
\end{abstract}
\maketitle

\section{Introduction}

General Relativity is a very successful physical theory in excellent agreement with experimental data. Gravitational waves in particular test strong and weak gravitational fields in the sense that a strong gravitational field is required for their creation while their propagation is governed by the weak field approximated field equations. Despite its success, General Relativity faces two observational challenges which are simply referred to as the dark matter and the dark energy problems. Moreover, from a theoretical point of view, quantum theory and gravity appear to be incompatible and no consensus exists yet regarding the form of such a theory. These issues have motivated the study of  early modifications of Einstein's theory like teleparallel gravity and Kaluza-Klein theories and, in the following years, have justified the proposal of a plethora of other theories.

One of the most studied and most tested of these models is the so-called scalar tensor theory of gravity.  These theories are based on the Brans-Dicke prototype action which is given by
\begin{align}
  S = \int \biggl\{ \frac{1}{2\kappa} \phi R - \frac{\omega}{\phi} \partial_a \phi \partial^a \phi \biggr\} \sqrt{-g}\, d^4x\,.
  \label{eq:bd1}
\end{align}
Here one includes a non-minimal coupling between the geometry in the form of the Ricci scalar $R$ and an additional scalar field $\phi$. $\omega$ is the Brans-Dicke parameter which in the limit $\omega \rightarrow \infty$ reduces Brans-Dicke theory to General Relativity. A fundamental problem related to this class of theories is connected to the nature of this scalar field. One hypothesis which is still under investigation is that this field might coincide with the Higgs field (see e.g.\cite{PhysRevD.53.6813,PhysRevD.62.023504,PALLIS2010287,1402-4896-82-1-015901,Calmet2018}). Another, perhaps more general, line of interpretation is that this scalar field is indeed an effective field representing a scalar degree of freedom of the theory. This happens, for example, in the scalar field representation of $f(R)$-gravity or the hybrid metric Palatini theories.

In this paper we present a new class of theories of gravitation in which the volume element is a dynamical object. In particular we wish to write $\sqrt{-\tilde{g}}:= \phi \sqrt{-g}$ and take $\sqrt{-\tilde{g}}\, d^4x$ as the fundamental volume form. On the level of the Brans-Dicke theory this means reinterpreting $\phi \sqrt{-g}\, d^4x$ as the volume element over which one has to integrate in order to evaluate the action. This idea can be associated by contrast to the so-called unimodular gravitational theories~\cite{Unruh:1988in,Buchmuller:1988wx,Buchmuller:1988yn,Finkelstein2000F,Shaposhnikov:2008jq,Ellis:2014iq}, in which the volume form is considered constant. In our case the volume form acts as an additional field: in the action, we will only change the volume element and retain the metric $g$ as the basic object from which curvature is computed.

At this stage there is no need to keep the simple relation between $\tilde{g}$ and $g$ which takes the form of a conformal coupling $\sqrt{-\tilde{g}}:= \phi \sqrt{-g}$. Instead, we will assume that $\tilde{g}$ is an arbitrary function of the metric $g$. Our key ingredient will be to assume that there exists a rank 4 tensor $\chi$ which relates those two metrics such that $\tilde{g}_{ab} = \chi_{ab}{}^{cd} g_{cd}$. This is in analogy to electromagnetism in materials where one has to distinguish between the electric field and the electric displacement and  relates to ideas discussed in~\cite{Boehmer:2013ss,Pearson:2014iaa,Carloni:2016glo,Boehmer:2014ipa}.  Clearly, this theory reduces to Brans-Dicke theory provided one takes a simple $\chi$ containing only Kronecker deltas and includes a kinetic term. As we will see, however this is not the only case in which our new theory can be shown to present only one additional degree of freedom. 

It is also worth pointing out an interesting link to the work in~\cite{Ellis:2010uc}: one can prove that these exist forms of $\chi$ for which  the gravitational field equations of the new theory become trace-free. This  result implies that we are dealing with a genuine generalisation of  unimodular gravity.

The paper is organised as following. Section II is  dedicated to the definition of the possible action(s) corresponding to the idea of a dynamical volume form and the derivation of their field equations and their properties. Section III, is instead dedicated to the exploration of the cosmology of the via phase space analysis of the most interesting of the actions defined in Section II. Section IV is dedicated to the conclusions.
 
\section{Model and gravitational field equations}

\subsection{Gravitational actions}

Following on from the previous discussion, we introduce the two actions
\begin{align}
  S_1 &= \int
  \biggl\{
  \frac{1}{2\kappa} R \sqrt{-\tilde{g}} + L_{\rm (m)}(g,\psi,\nabla\psi) \sqrt{-g}
  \biggr\} d^4x\,, 
  \label{eq:n1} \\
  S_2 &= \int
  \biggl\{
  \frac{1}{2\kappa} R \sqrt{-\tilde{g}} + L_{\rm (m)}(\tilde{g},\psi,\nabla\psi) \sqrt{-\tilde{g}}
  \biggr\} d^4x\,.
  \label{eq:n2}
\end{align}
The main difference between those two actions is the coupling of the matter in the theory, clearly the most important issue when it comes to any gravitational theory. Since there is little guidance as to which of those two is preferred from a theoretical point of view, we will study both cases separately. As one can probably expect at this point, these two versions of the theory will give rise to quite a different phenomenology when applied to cosmology. Ideally, some external input like observational data could be used to make this choice. As we will see, it turns out that the most interesting cosmological models are given by $S_1$.

In the following we will assume the relationship \begin{align}
  \tilde{g}_{ab} = \chi_{ab}{}^{cd} g_{cd} \,,
  \label{gtilde}
\end{align}
between the two metrics. As we will see, this includes a number of interesting cases e.g.~conformal/disformal transformations which were studied in different contexts. Note that the standard Brans-Dicke theory is recovered when $\chi_{ab}{}^{cd} = \delta_a^c \delta_b^d \phi^{1/4}$.

The volume integration in $S_2$ is straightforward in the sense that one views $\sqrt{-\tilde{g}} d^4x$ as the volume form of spacetime while keeping in mind that $R$ is computed using the metric $g$. In this case we require $\tilde{g}$ to be a well-defined metric which means the inverse of $\tilde{g}$ must also exist. Hence one arrives at
\begin{align}
  \tilde{g}_{km}\tilde{g}^{mn} = \chi_{km}{}^{cd} g_{cd} \tilde{g}^{mn} =
  \chi_{km}{}^{ab} g_{ab} \chi^{mn}{}_{cd} g^{cd} = \chi_{kma}{}^{a} \chi^{mnc}{}_{c} = \delta_k^n\,.
  \label{gtilde2}
\end{align}
This means we impose the following condition on the tensor $\chi_{km}{}^{cd}$
\begin{align}
  \chi_{kma}{}^{a} \chi^{mnc}{}_{c} = \delta_k^n\,,
  \label{gtilde2b}
\end{align}
where the metric $g$ was used to raise and lower indices. 

On the other hand, in $S_1$ one is using two different volume elements and one might be tempted to regard this as unnatural. However, one can rewrite $S_1$ as follows
\begin{align}
  S_1 &= \int
  \biggl\{
  \frac{1}{2\kappa} R \frac{\sqrt{-\tilde{g}}}{\sqrt{-g}} + L_{\rm (m)}(g,\psi,\nabla\psi)
  \biggr\} \sqrt{-g} d^4x =
  \int
  \biggl\{
  \frac{1}{2\kappa} R\, \rho + L_{\rm (m)}(g,\psi,\nabla\psi)
  \biggr\} \sqrt{-g} d^4x \,,
  \label{eq:n1a}
\end{align}
where we introduced $\rho = \sqrt{-\tilde{g}}/\sqrt{-g}$, thereby integrating over an appropriate volume\footnote{Note that the determinant of a metric is not a true scalar under arbitrary coordinate transformations. However, we will find in Sec.~\ref{sec:GFE} that $\rho$ is a true scalar field.}. The possibility of writing the action might suggest that theories of the type $S_1$ and $S_2$ contain only an additional degree of freedom. This can appear strange, as the tensor $\chi$ has great number of non trivial components. Indeed we will find that for a surprisingly general form of $\chi$ these theories present only one additional degree of freedom

One can view the matter coupling in $S_1$ as minimal since the matter couples to gravity via the canonical volume form. Likewise, one can view the matter coupling in $S_2$ as non-minimal because the matter couples to gravity via $\tilde{g}$ which in itself obeys a relation to $g$. When stating the field equations explicitly, this point will be verified. 

\subsection{Determinant and variations of $\tilde{g}$}

The determinant of a rank 2 tensor $M_{ij}$ is defined by
\begin{align}\label{Det_gen}
  M = \det(M_{ij}) = 
  \frac{1}{4!} \varepsilon^{ijkl} \varepsilon^{abcd} M_{ia} M_{jb} M_{kc} M_{ld} \,,
\end{align}
where we work using the standard convention $\varepsilon_{0123}=1$. When raising or lowering indices of $\varepsilon_{ijkl}$, one has to be quite careful and  work with the (pseudo) tensor  
\begin{align}
  \eta_{ijkl} = \sqrt{-g} \varepsilon_{ijkl}\,, \quad
  \eta^{ijkl} = -\frac{\varepsilon^{ijkl}}{\sqrt{-g}} \,.
\end{align}
Now, we can define the determinant of $\tilde{g}_{ij}$ which gives
\begin{align}
  \tilde{g} &= \det(\tilde{g}_{ij}) = 
  \frac{1}{4!} \varepsilon^{ijkl} \varepsilon^{abcd}
  \tilde{g}_{ia} \tilde{g}_{jb} \tilde{g}_{kc} \tilde{g}_{ld}
  \nonumber \\ &=
  \frac{1}{4!} \varepsilon^{ijkl} \varepsilon^{abcd}
  \chi_{ia}{}^{pq} g_{pq} \chi_{jb}{}^{rs} g_{rs} \chi_{kc}{}^{tu} g_{tu} \chi_{ld}{}^{vw} g_{vw}
  \nonumber \\ &=
  \frac{1}{4!} \Bigl(\varepsilon^{ijkl} \varepsilon^{abcd} \chi_{ia}{}^{pq} \chi_{jb}{}^{rs} \chi_{kc}{}^{tu} \chi_{ld}{}^{vw}\Bigr)  g_{pq} g_{rs} g_{tu} g_{vw}\,.
\end{align}
The above formula can be used to calculate the variations of $\tilde{g}$ with respect to the metric. We begin with
\begin{align}
  \delta \tilde{g} &=
  \frac{1}{4!} \Bigl(\varepsilon^{ijkl} \varepsilon^{abcd} \chi_{ia}{}^{pq} \chi_{jb}{}^{rs} \chi_{kc}{}^{tu} \chi_{ld}{}^{vw}\Bigr) (\delta g_{pq} g_{rs} g_{tu} g_{vw} + g_{pq} \delta g_{rs} g_{tu} g_{vw} + g_{pq} g_{rs} \delta g_{tu} g_{vw} + g_{pq} g_{rs} g_{tu} \delta g_{vw})
  \nonumber \\ &=
  \frac{1}{4!} \Bigl(\varepsilon^{ijkl} \varepsilon^{abcd} \chi_{ia}{}^{pq} \chi_{jb}{}^{rs} \chi_{kc}{}^{tu} \chi_{ld}{}^{vw}\Bigr) (\delta^m_p \delta^n_q g_{rs} g_{tu} g_{vw} \nonumber \\
  &~~~+ g_{pq} \delta^m_r \delta^n_s g_{tu} g_{vw} + g_{pq} g_{rs} \delta^m_t \delta^n_u g_{vw} + g_{pq} g_{rs} g_{tu} \delta^m_v \delta^n_w) \delta g_{mn}
\end{align}
Let us use the standard identity $\delta g_{mn} = - g_{mi} g_{nj} \delta g^{ij}$ and the implicit definition for $\overline{\chi}_{mn}$ given by
\begin{align}
  \delta \tilde{g} &=: - \tilde{g} \, \overline{\chi}_{mn}\, \delta g^{mn} \,.
  \label{eqn:varigtilde}
\end{align}
which is motivated by the GR analogue of this equation. Then, after some algebra, we arrive at
\begin{multline}
  \overline{\chi}_{mn} = -\frac{1}{4!} \Bigl(\eta^{ijkl} \eta^{abcd} \chi_{ia}{}^{pq} \chi_{jb}{}^{rs} \chi_{kc}{}^{tu} \chi_{ld}{}^{vw}\Bigr) \\ \Bigl[ g_{pm} g_{qn} g_{rs} g_{tu} g_{vw} + g_{pq} g_{rm} g_{sn} g_{tu} g_{vw} + g_{pq} g_{rs} g_{tm} g_{un} g_{vw} + g_{pq} g_{rs} g_{tu} g_{vm} g_{wn} \Bigr] \,.
  \label{eq:chibardef}
\end{multline}
The quantity $\overline{\chi}_{mn}$ will enter the equations of motion or field equations through the variations of the volume element with respect to the metric. Note also that from the general expression \eqref{Det_gen} one has
\begin{align}
  \overline{\chi}_{m}{}^m= -\frac{1}{3!} \Bigl(\eta^{ijkl} \eta^{abcd} \tilde{g}_{ia} \tilde{g}_{jb}{}\tilde{g}_{kc}{}\tilde{g}_{ld}{}\Bigr) = 4\frac{\tilde{g}}{g} = 4 \rho^2.
\end{align}
The factor $4$ appears because the trace of the tensor $\overline{\chi}$ corresponds to the trace of the Kronecker delta when $g$ and $\tilde{g}$ coincide, which also gives $\rho=1$.

\subsection{Gravitational equations of motion} \label{sec:GFE}

Having established the variations of the new volume form, we are now ready to derive the equations of motion of the gravitational part of the action. Matter couplings will be addressed separately in the next Section. Let us begin with
\begin{align}
  \delta\Bigl[ g^{ab} R_{ab} \sqrt{-\tilde{g}} \Bigr] = 
  \delta g^{ab} R_{ab} \sqrt{-\tilde{g}} + 
  g^{ab} \delta R_{ab} \sqrt{-\tilde{g}} -
  \frac{1}{2}\frac{1}{\sqrt{-\tilde{g}}} g^{ab} R_{ab} \delta \tilde{g} \,.
\end{align}
Using the above equation (\ref{eqn:varigtilde}) for the variation of $\tilde{g}$, we have
\begin{align}
  \delta\Bigl[ g^{ab} R_{ab} \sqrt{-\tilde{g}} \Bigr] = 
  \Bigl[R_{ab} - \frac{1}{2}R\, \overline{\chi}_{ab} \Bigr] \sqrt{-\tilde{g}} \delta g^{ab} +
  g^{ab} \sqrt{-\tilde{g}}\, \delta R_{ab}\,.
  \label{eqn:vari_R}
\end{align}
In GR this final term would be a boundary term, however, for the current model this is not the case. We recall
\begin{align}
  \delta R_{mn} = \nabla_d \delta \Gamma^d_{mn} - \nabla_m \delta \Gamma^d_{dn} \,.
\end{align}
and use the formula
\begin{align}
  g^{mn} \sqrt{-\tilde{g}}\, \delta R_{mn} &= \frac{1}{2} g^{mn}\left(2\nabla^{a}\nabla_{(m}\delta g_{n) a}-g^{ab}\nabla_{m}\nabla_{n}\delta g_{a b}- \nabla_{a}\nabla^{a}\delta g_{mn} \right)\sqrt{-\tilde{g}}\nonumber\\
  &=\left(\nabla^{a}\nabla^{b}\delta g_{a b}-g^{mn} \nabla_{a}\nabla^{a}\delta g_{mn}\right) \sqrt{-\tilde{g}}
\end{align}
In this way, recalling $\rho = \sqrt{-\tilde{g}}/\sqrt{-g}$ to eliminate the boundary terms, we have
\begin{align}
g^{mn} \sqrt{-\tilde{g}}\, \delta R_{mn} = \left(\nabla^{a}\nabla^{b}\rho-g^{ab} \nabla_{c}\nabla^{c}\rho\right) \sqrt{-{g}} \, \delta g_{a b} \,.
\end{align}
Note that $\rho$ is a true scalar in the sense of differential geometry. While determinants are pseudo-scalars (they transform differently to scalars), the ratio of two pseudo-scalar gives a scalar field as stated previously. Consequently, we see that this ratio of `volumes' looks like the Brans-Dicke scalar. Putting everything together leads to
\begin{align}
  \delta\Bigl[ g^{ab} R_{ab} \sqrt{-\tilde{g}} \Bigr] = 
  \Bigl[\rho\left(R_{ab} - \frac{1}{2}R\, \overline{\chi}_{ab}\right) -\nabla_{a}\nabla_{b}\rho+g_{ab} \nabla_{m}\nabla^{m}\rho\Bigr] \sqrt{-g}\, \delta g^{ab} \,.
  \label{eqn:vari_R2}
\end{align}
Next, we need to couple matter to the geometry.

\subsection{Coupling matter with $\sqrt{-g}$}

The minimal coupling setting is described by action $S_1$, given by (\ref{eq:n1}). A direct calculation gives
\begin{align}
  \delta\Bigl[  L_{\rm (m)}(g,\psi,\nabla\psi)\sqrt{-g} \Bigr] = 
  T_{ab} \sqrt{-g} \delta g^{ab} \,,
\end{align}
so that the complete field equations take the form
\begin{align}
  \rho G_{ab} &= 2\kappa T_{ab} + \frac{1}{2}\rho R \left( \bar{\chi}_{ab}-g_{ab}\right)+\nabla_{a}\nabla_{b}\rho-g_{ab} \nabla_{m}\nabla^{m}\rho \,,
  \label{eq:Ein_n2b_2}
\end{align}

Should one consider the issue of a field equation for the field $\rho$? As $\rho$ is not a dynamical variable for which we specify a separate Lagrangian, one would not expect it to satisfy additional field equations. However, general relativity and its modifications obey other symmetry properties so that one cannot simply choose $\rho$ freely. To see this, consider the trace of the (\ref{eq:Ein_n2b_2}) which gives
\begin{align}
  3\Box\rho+\rho R - \frac{1}{2}\rho \bar{\chi}R - 2\kappa T=0\,,
\end{align}
which has the structure of a Klein-Gordon type equation for $\rho$. Due to the presence of $\bar{\chi}_{ab}$ in Eq.~(\ref{eq:Ein_n2b_2}) this theory is distinct from scalar field theories.

\subsection{Trace free field equations}

Notice that Eq.~(\ref{eq:Ein_n2b_2}) can give the trace free equations. In fact, we can write (\ref{eq:Ein_n2b_2}) as
\begin{align}
  \rho \left(R_{ab} - \frac{1}{2} R \bar{\chi}_{ab}\right) &= \nabla_{a}\nabla_{b}\rho-g_{ab} \nabla_{m}\nabla^{m}\rho+ 2\kappa T_{ab}\,.
\end{align}
The left-hand side of the latter equation becomes the trace-free equation if we make the choice $\bar{\chi}_{ab}=g_{ab}/2$. To achieve this, let us choose for instance $\chi_{ab}{}^{cd} = \Omega \delta_a^c \delta_b^d$, then
\begin{align}
  \tilde{g}_{ab} = \chi_{ab}{}^{cd} g_{cd} = \Omega \delta_a^c \delta_b^d g_{cd} = 
  \Omega g_{ab} \,.
  \label{iso4}
\end{align}
We begin with the first term of (\ref{eq:chibardef}) and get
\begin{align}
  \Bigl(\eta^{ijkl} \eta^{abcd} \chi_{ia}{}^{pq} \chi_{jb}{}^{rs} \chi_{kc}{}^{tu} \chi_{ld}{}^{vw}\Bigr) =
  \Omega^4 \Bigl(\eta^{ijkl} \eta^{abcd} \delta_i^p \delta_a^q \delta_j^r \delta_b^s 
  \delta_k^t \delta_c^u \delta_l^v \delta_d^w \Bigr) = 
  \Omega^4 \eta^{prtv} \eta^{qsuw} \,,
\end{align}
and consequently one arrives at
\begin{align}
  \overline{\chi}_{mn} = \Omega^4 g_{mn} \,.
\end{align}
This means $\Omega^4 =1/2$ gives a trace free or conformal left-hand side. Note that in this case
\begin{align}
  \det(\tilde{g}_{ab}) = \Omega^4 \det(g_{ab}) \quad \Rightarrow \quad
  \rho^2 = \frac{\det(\tilde{g}_{ab})}{\det(g_{ab})} = \Omega^4
  \quad \Rightarrow \quad 
  \rho^2 = \Omega^4 = 1/2\,.
\end{align}
This is an elegant result which fits into the framework considered by Ellis \cite{Ellis:2010uc,Ellis:2014iq}.

\subsection{Coupling matter with $\sqrt{-\tilde{g}}$}

Considering action (\ref{eq:n2}) the variation of the matter part would introduce terms that contain explicitly the matter Lagrangian
\begin{align}
  \delta\Bigl[ L_{\rm (m)}(g,\psi,\nabla\psi)\sqrt{-\tilde{g}} \Bigr] = 
  \Bigl[\rho T_{ab}+  \frac{1}{2}\rho \left( g_{ab}-\bar{\chi}_{ab}\right) L_{\rm (m)}\Bigr] 
  \sqrt{-g} \delta g^{ab} \,.
\end{align}
This leads to the following set of field equations
\begin{align}
  \rho G_{ab} &= 2 \rho \kappa T_{ab} + \frac{1}{2}\rho \left[R+2\kappa L_{\rm (m)}\right]
  \left( \bar{\chi}_{ab}-g_{ab}\right)+\nabla_{a}\nabla_{b}\rho-g_{ab} \nabla_{m}\nabla^{m}\rho\,. 
  \label{eq:Ein_n2b_A}
\end{align}
The explicit dependence of the field equations on the matter Lagrangian means that we are dealing with a non-minimally coupled theory. While theories of this type have been considered in the past, such models are problematic. We should note that when setting $\kappa L_{\rm (m)} = 0$ in the field equation (\ref{eq:Ein_n2b_A}), one recovers the previous field equation (\ref{eq:Ein_n2b_2}) in vacuum, which is perhaps unsurprising as the respective actions only differ by the form of the matter coupling to geometry. This implies that the two actions present, for example, the same black hole solutions.

\subsection{The limit to General Relativity}
\label{sec:GRconnections}

Starting with Eq.~(\ref{gtilde}), let us recall the most general isotropic rank 4 tensor which is given by
\begin{align}
  \chi_{ab}{}^{cd} = \alpha g_{ab} g^{cd} + 
  \beta \delta_a^c \delta_b^d + \gamma \delta_a^d \delta_b^c \,,
  \label{iso}
\end{align}
where $\alpha$, $\beta$ and $\gamma$ are some functions of the coordinates in general. However, we assume those to be constants to recover General Relativity. This gives
\begin{align}
  \tilde{g}_{ab} = \chi_{ab}{}^{cd} g_{cd} = (\alpha g_{ab} g^{cd} + 
  \beta \delta_a^c \delta_b^d + \gamma \delta_a^d \delta_b^c) g_{cd} =
  4\alpha g_{ab} + \beta g_{ab} + \gamma g_{ba} = (4\alpha+\beta+\gamma) g_{ab} \,.
  \label{iso2}
\end{align}
Let us introduce the notation $\Omega^2 = 4\alpha + \beta + \gamma$. This is a natural choice because (\ref{iso2}) should preserve the metric signature, so that this result can now be written in the familiar looking form
\begin{align}
  \tilde{g}_{ab} = \Omega^2 g_{ab} \,.
  \label{iso3}
\end{align}
This takes the form of a conformal transformation. Choosing $\alpha,\beta,\gamma$ such that $\Omega=1$ reduce this theory to General Relativity. For simplicity, we choose $\alpha=\gamma=0$ so that $\chi_{ab}{}^{cd} = \delta_a^c \delta_b^d$ and hence $\tilde{g}_{ab} = \delta_a^c \delta_b^d g_{cd} = g_{ab}$. This also implies that $\det(\tilde{g}_{ab}) = \det(g_{ab})$ and therefore $\rho = 1$.

It remains to check that $\overline{\chi}_{mn}=g_{mn}$ given this specific choice of $\chi$. We begin with the first term of Eq.~(\ref{eq:chibardef}) and find
\begin{align}
  \Bigl(\eta^{ijkl} \eta^{abcd} \chi_{ia}{}^{pq} \chi_{jb}{}^{rs} \chi_{kc}{}^{tu} \chi_{ld}{}^{vw}\Bigr) =
  \Bigl(\eta^{ijkl} \eta^{abcd} \delta_i^p \delta_a^q \delta_j^r \delta_b^s 
  \delta_k^t \delta_c^u \delta_l^v \delta_d^w \Bigr) = 
  \eta^{prtv} \eta^{qsuw} \,.
\end{align}
Consequently, one arrives at
\begin{align}
  \overline{\chi}_{mn} = -\frac{1}{4!}
  \eta^{prtv} \eta^{qsuw}\Bigl[ g_{pm} g_{qn} g_{rs} g_{tu} g_{vw} + g_{pq} g_{rm} g_{sn} g_{tu} g_{vw} + g_{pq} g_{rs} g_{tm} g_{un} g_{vw} + g_{pq} g_{rs} g_{tu} g_{vm} g_{wn} \Bigr] \,.
\end{align}
Each of these four terms will contribute in identical ways. To see this note
\begin{align}
  \eta^{prtv} \eta^{qsuw}g_{pm} g_{qn} g_{rs} g_{tu} g_{vw} &= 
  (\eta^{prtv} \eta^{qsuw}g_{pm} g_{qn} g_{rs}) g_{tu} g_{vw}\\
  \eta^{prtv} \eta^{qsuw}g_{pq} g_{rm} g_{sn} g_{tu} g_{vw} &=
  (\eta^{prtv} \eta^{qsuw}g_{pq} g_{rm} g_{sn}) g_{tu} g_{vw} =
  (\eta^{prtv} \eta^{qsuw}g_{rm} g_{sn} g_{pq}) g_{tu} g_{vw} 
  \nonumber \\&=
  (\eta^{rptv} \eta^{squw}g_{pm} g_{qn} g_{rs}) g_{tu} g_{vw} =
  ((-)\eta^{prtv} (-)\eta^{qsuw}g_{pm} g_{qn} g_{rs}) g_{tu} g_{vw} 
  \nonumber \\&=
  (\eta^{prtv} \eta^{qsuw}g_{pm} g_{qn} g_{rs}) g_{tu} g_{vw} \,.
\end{align}
So, we see that the first and second term match. Therefore
\begin{align}
  \overline{\chi}_{mn} = -\frac{1}{3!}
  \eta^{prtv} \eta^{qsuw}\Bigl[ g_{pm} g_{qn} g_{rs} g_{tu} g_{vw}\Bigr] =
  -\frac{1}{3!}
  \Bigl[\eta^{prtv} \eta^{qsuw} g_{rs} g_{tu} g_{vw}\Bigr] g_{pm} g_{qn} \,.
\end{align}
Using the standard identity
\begin{align}
  \Bigl[\eta^{prtv} \eta^{qsuw} g_{rs} g_{tu} g_{vw}\Bigr] = -3! g^{pq}\,,
\end{align}
yields the desired equation
\begin{align}
  \overline{\chi}_{mn} = -\frac{1}{3!}
  (-3!) g^{pq} g_{pm} g_{qn} = g_{mn} \,.
\end{align}
In summary the GR limit means: $\tilde{g}_{ab} = g_{ab}$, $\overline{\chi}_{mn} = g_{mn}$ and $\rho=1$.

Substituting these back into the two sets of field equations, one can easily verify that both \eqref{eq:Ein_n2b_2} and \eqref{eq:Ein_n2b_A} reduce to the Einstein field equations.

\section{Applications to cosmology}

In the following we will investigate the cosmological dynamics of Model 1 which gives rise to various interesting properties. For completeness, we also studied Model 2 which is discussed in Appendix~\ref{app:model2}. Model 1 allows for solutions which make a transition from acceleration to deceleration and then back to acceleration before terminating at a scaling solution. This feature makes this model particularly interesting when applied to our Universe. 

\subsection{Choosing a cosmological $\chi$}

Let us consider \eqref{eq:Ein_n2b_2} and \eqref{eq:Ein_n2b_A} in the case of cosmological spacetimes.  In spite of the compact form of the field equations that we have given using the quantity $\rho$, one should not forget that these equations depend on all the components of the tensor $\chi$. Treating the most general form of these equations is a formidable task which we will not undertake here. Rather we will consider a general case in which the theory can be considered as function only of the metric tensor and $\rho$. This restriction on the degrees of freedom of the theory allows a connection with Brans Dicke theory. As we will see, however, this resemblance is only apparent, as \eqref{eq:Ein_n2b_2} and \eqref{eq:Ein_n2b_A} present interesting peculiarities.

Starting form the general form \eqref{iso}, we will assume that the tensor $\chi_{abcd}$ can be written as 
\begin{align}
\chi_{abcd}= \chi_1 u_a u_b u_c u_d + 2\chi_2 u_a u_{(c} h_{d)b}+ 2\chi_3 h_{a (c}u_{d)} u_b+ \chi_4 h_{ac}h_{b d}+ \chi_5 u_a u_b h_{cd}+ \chi_6 h_{ab}u_c u_d+ \chi_7 h_{ab}h_{c d} \,,
\end{align}
where $u_a$ is the velocity of a chosen observer,  $h_{ab}=g_{ab}-u_au_b$ and $\chi_i$ are generic functions of the cosmic time. Such a  choice however would not lead to a set of equations containing only the metric and $\rho$. In order to obtain equations of this type in a fairly general way we can set  
\begin{align}
 \chi_5 \rightarrow \alpha \chi_1 \quad  \chi_6 \rightarrow  \beta \chi_4\quad  \chi_7 \rightarrow  \gamma \chi_4 \,,
\end{align}
so that 
\begin{align}\label{Chi_ansatz_1}
\chi_{abcd}= \chi_1 u_a u_b (u_c u_d +\alpha h_{cd}) + 2\chi_2 u_a u_{(c} h_{d)b}+ 2\chi_3 h_{a (c}u_{d)} u_b++ \chi_4 [h_{ac}h_{b d}+ \gamma(  \beta u_c u_d + h_{cd})] \,.
\end{align}
In the following we will use this form of $\chi_{abcd}$ in the context of cosmology.

\subsection{The cosmology of action $S_1$}

Using the standard FLRW metric and the previous choice \eqref{Chi_ansatz_1} we find
\begin{align}\label{SysCosm1}
  \left(2\rho ^2-1\right)\left(\dot{H}+H^2\right) &=
  -\frac{2\kappa  \mu }{3\rho } \left\{1-[2+\mathbf{A}(3w-1)] \rho ^2 \right\}+
  \frac{H \dot{\rho} }{2 \rho }\left[(3\mathbf{A}-2) \rho ^2+1\right]+\mathbf{A}\ddot{\rho} \rho\,, \\
  \left(2\rho ^2-1\right)\left(H^2+\frac{k}{S ^{2}}\right) &=
  \frac{\kappa  \mu }{3 \rho } \left\{ 2[2+\mathbf{A}(3w-1)] \rho ^2-(1+3w)\right\}
  +\frac{H \dot{\rho}}{2 \rho  }\left[2(3\mathbf{A}-2) \rho ^2-1\right] +\frac{ \ddot{\rho}}{2 \rho  }\left( 2 \mathbf{A}\rho ^2-1\right)\,,\\
  \dot{\mu}+3H (\mu+p) &= 0 \,,
\end{align}
and the final equation
\begin{multline}
  2\rho\left(2\rho ^2-1\right)\left(2 \mathbf{A}\rho ^2-1\right)\dddot{\rho} +
  \left[H B_1(\rho,\mathbf{A})+B_2(\rho,\mathbf{A})\dot{\rho}\right]\ddot{\rho} +
  H B_3(\rho,\mathbf{A})  \dot\rho^2  \\ +
  \left\{2 \kappa  \mu  \left[B_4(\rho,\mathbf{A})+wB_5(\rho,\mathbf{A})\right]+
  \frac{2 k }{S^2} B_6(\rho,\mathbf{A}) + H^2 B_7(\rho,\mathbf{A}) \right\}\dot{\rho} \\ =
  2 H \kappa \mu  \left[B_8(\rho,\mathbf{A})-w B_9(\rho,\mathbf{A})+w^2 B_{10}(\rho,\mathbf{A}) \right] \,.
\end{multline}
Here we introduced the following functions
\begin{align}
  \mathbf{A}&=\frac{ \gamma  (9 \alpha  \beta -4 \beta +3)+1}{(3 \alpha
    -1) ((\beta -3) \gamma -1)} \,, \\[1ex]
  B_1(\rho,\mathbf{A})&=  \rho  \left(2 \rho ^2-1\right) \left[2 (7 \mathbf{A}-2) \rho ^2-3\right] \,, \\
  B_2(\rho,\mathbf{A})&=   -16 \mathbf{A} (6 \mathbf{A}^2-7\mathbf{A}+2) \rho ^6+8 (5 \mathbf{A}^2-5\mathbf{A}+2) \rho ^4+(6 \mathbf{A}-20) \rho ^2+5 \,, \\
  B_3(\rho,\mathbf{A})&=8 \left(3 \mathbf{A}^2+\mathbf{A}-2\right) \rho ^4-16 (2-3 \mathbf{A})^2 (2 \mathbf{A}-1) \rho ^6+10
  (\mathbf{A}-2) \rho ^2+3 \,, \\
  B_4(\rho,\mathbf{A})&= \frac{2}{3} \rho ^2 \left(8 (\mathbf{A}-2) (2 \mathbf{A}-1) (3 \mathbf{A}-2) \rho ^4+4 (7 \mathbf{A}^2-11\mathbf{A}+6) \rho ^2+\mathbf{A}-10\right)+\frac{7}{3} \,, \\
  B_5(\rho,\mathbf{A})&=-16 \mathbf{A}  (6 \mathbf{A}^2-7\mathbf{A}+2) \rho ^6+8 (5 (\mathbf{A}-1) \mathbf{A}+2) \rho ^4+(6 \mathbf{A}-20) \rho ^2+5 \,, \\
  B_6(\rho,\mathbf{A})&=\rho\left(2 \rho ^2-1\right) \left[8 (6 \mathbf{A}^2-7\mathbf{A}+2) \rho ^4+2 (5 \mathbf{A}-4) \rho^2+3\right] \,, \\
  B_7(\rho,\mathbf{A})&= 8 (2 \mathbf{A}-1)  \rho ^3 \left(2 \rho ^2-1\right) \left[(6 \mathbf{A}-4) \rho ^2-1\right] \,, \\
  B_8(\rho,\mathbf{A})&=\frac{4}{3} \rho  \left(2 \rho ^2-1\right) \left(2 (\mathbf{A}-2) \rho ^2+3\right) \,, \\
  B_9(\rho,\mathbf{A})&=8 \rho  \left(2 \rho^2-1\right)  \left(2 (\mathbf{A}-1) \rho ^2+1\right) \,, \\
  B_{10}(\rho,\mathbf{A})&=12 \rho \left(2 \rho ^2-1\right)  \left(2 \mathbf{A} \rho ^2-1\right)\,.
\end{align}

We can easily explore this cosmological model via phase space analysis, see \cite{Coley:1999uh,Coley:2003mj,Boehmer:2014vea,Bahamonde:2017ize}, setting 
\begin{align}
X=\frac{\dot\rho}{4H\rho} \,,\qquad Y=\frac{\ddot\rho}{4H^2\rho}\,, \qquad Z=\rho \,,\qquad \Omega = \frac{\kappa \mu}{3 \rho H^2} \,,\qquad K=\frac{k}{ S^2 H^2} \,.
\end{align}

Choosing the time variable $\tau=\ln (S/S_0)$ the dynamical equations are
\begin{align}\label{DynSys1}
\begin{split}
X' &=\frac{1 }{4 \mathbf{A} Z^2-2}\bigg\{K\left[Z^2 (4 \mathbf{A} X+2)-1\right]-12 \mathbf{A} w
   \Omega  Z^2-16 X^2 \left(\mathbf{A} Z^2-1\right)\\
   &~~~-4 X \left[(\mathbf{A}-2) Z^2+2 \Omega \right]+
   4 \Omega Z^2\left[ \left( \mathbf{A}  -2 \right) + 2\left(1+3 w \right)\right]
  +2 Z^2-1 \bigg\}\,,\\
Z' &= 8 X Z\,,\\
 K'&=\frac{4 K }{2 \mathbf{A} Z^2-1}\left[\mathbf{A} (K+1) Z^2+2 X-2 \Omega \right]\,,\\
 \Omega' &=\frac{\Omega }{2 \mathbf{A} Z^2-1} \left[4 \mathbf{A} K Z^2+w \left(3-6 \mathbf{A} Z^2\right)+X \left(12-8 \mathbf{A} Z^2\right)+2
   \mathbf{A} Z^2-8 \Omega +1\right]\,.
 \end{split}
\end{align}
where the prime denotes the derivative with respect to $\tau$ and we applied the constraint
\begin{align}
\Omega  \left\{2 Z^2 [(1-3w)\mathbf{A}-2]+3 w+1\right\}+X \left[(4-6 \mathbf{A}) Z^2+1\right]+Y \left(1-2
   \mathbf{A} Z^2\right)+\frac{1}{2} (K+1) \left(2 Z^2-1\right)=0 \,.
\end{align}
The solutions associated to the fixed points can be found solving the equation
\begin{align}
  H'= \frac{H}{1-2 \mathbf{A} Z_*^2}
  \left(2 \mathbf{A} K_* Z_*^2+4 \mathbf{A} Z_*^2+4 X_*-4 \Omega_* -1\right) \,,
\end{align}
where an asterisk denotes the value of the variables in the fixed point. The fixed points an their stability can be found in Table \ref{Tab:EqA_1}. 

\begin{table}
\begin{center}
\begin{tabular}{ccccc}
 Point & Coordinates & Attractor & Repeller & Solutions \\\hline
${\mathsf L}$ & $\{K\to -1,\Omega \to 0,X\to 0, Z\to Z_0\}$ & $Z_0>0$ & Never & $a\to a_0 \left(t-t_0\right)$ \\ \\
$\mathrm{A}$  & $\left\{K\to -1,\Omega \to 0,X\to 0 ,Z\to 0\right\}$ & Never & Never & $a\to
   a_0 \left(t-t_0\right)$ \\ \\
$\mathsf{B}$  & $\left\{K\to 0,\Omega \to 0,X\to -\frac{1}{4} ,Z\to 0\right\}$ & Never & Never & $a\to a_0 \sqrt{2 t-t_0}$ \\ \\
$\mathsf{C}$  & $\left\{K\to 0,\Omega \to 0,X\to \frac{1}{4},Z\to 0\right\}$ & Never & Never & $a\to a_0 e^{H_0 t}$ \\ \\
$\mathsf{D}$  & $\{K\to 0,\Omega \to 0,X\to 0,Z\to \frac{1}{\sqrt{2}}\}$ & $0<\mathbf{A}<1$& Never & $a\to a_0 \left(t-t_0\right)^{\frac{\mathbf{A}-1}{2 \mathbf{A}-1}}$\\ \\
$\mathsf{E}$  & $\left\{K\to 0,\Omega \to \frac{1}{4}(2-3w),X\to \frac{1}{4}(1-3w),Z\to 0\right\}$ & Never & Never & $a\to a_0 \sqrt{2 t-t_0}$ \\ \\
$\mathsf{F}$  & $\left\{K\to 0,\Omega \to \frac{1+3 w-(3w-1)\mathbf{A}}{4 (3 \mathbf{A} w-\mathbf{A}+2)},X\to
   0,Z\to \sqrt{\frac{3 (w+1)}{2[(3w-1)\mathbf{A}+2]}}\right\}$ & Never & Never & $a\to a_0 \left(t-t_0\right)^{\frac{2}{3(1+w)}}$ \\ \\ \hline
\end{tabular}
\end{center}
\caption{Critical points stability and associated solution of Equations A with the ansatz \eqref{Chi_ansatz_1}.}
\label{Tab:EqA_1}
\end{table}

Since $\rho>0$ by definition, only the $Z>0$ part of the phase space has physical meaning. We will refer to this part of the phase space as {\it physical}. The system \eqref{DynSys1} contains three invariant submanifolds $\Omega=0$, $K=0$ and $Z=0$. Therefore a global attractor  for the cosmology has to lay in the intersection of these three submanifolds. The system also presents a singular submanifold $Z=(2A)^{-1/2}$, which is related to the structure of \eqref{SysCosm1} and in particular to the factor multiplying the left hand side of the Friedmann and Raychaudhuri equations. The physical phase space contains a line of fixed points together with six isolated points of which one belongs to this line. The phase space presents two different attractors: the Line~$\mathsf{L}$ and Point~$\mathsf{D}$, none of them global. Of these, only the stability properties of Point~$\mathsf{D}$ depend on the value of $\mathbf{A}$. The presence of these attractors suggests that, depending on initial conditions, the final state of the cosmology might be very different.  The phase space also presents a saddle point which corresponds to a de Sitter solution, leaving space for a transient phase of accelerated expansion. 

\begin{figure}
\centering
\includegraphics[width=0.8\columnwidth]{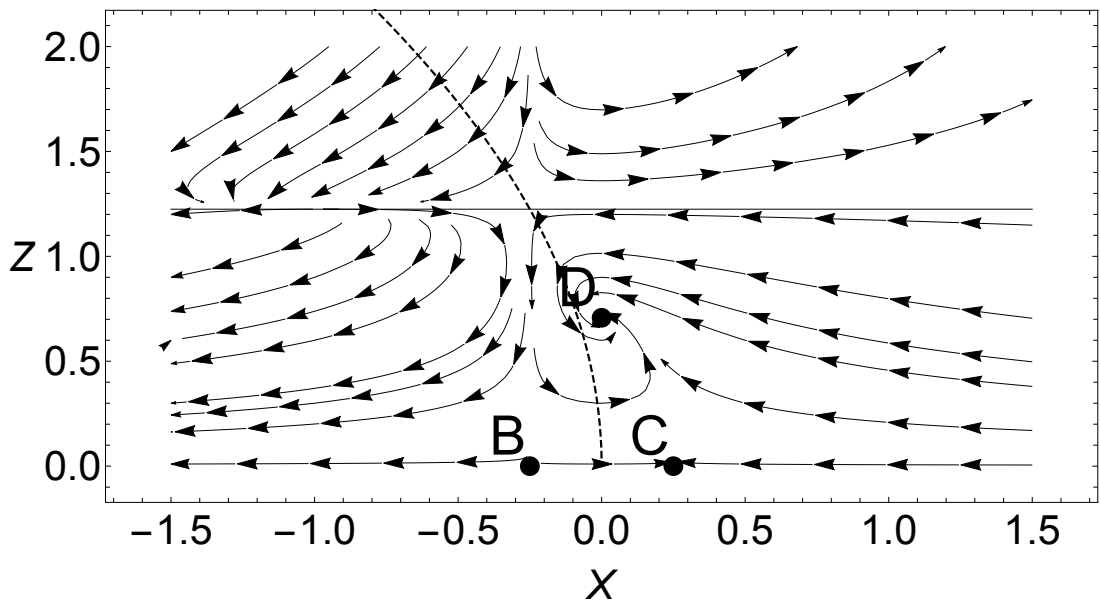}
\caption{The phase space for Equations A with the ansatz \eqref{Chi_ansatz_1} for spatially flat spacetime and vacuum. Here $\mathbf A =1/3$ and the dashed line separates accelerating (right part of the plot) from decelerating expansion (left part of the plot).}
\label{Fig:EqA_1}
\end{figure}

To gain an idea of the dynamics, we can look at the phase space of this model in the spatially flat ($K=0$) and vacuum ($\Omega=0$) case (see Fig.~\ref{Fig:EqA_1} in which $\mathbf{A}=1/3$). It is evident that four different types of cosmic histories are possible depending on the initial conditions. The most interesting is the one when $X>-1/4$ and $Z<\sqrt{3/2})$ in which the Universe can have initial conditions in accelerated expansion then switches to decelerated expansion $(\mathsf{B})$ to accelerate again first with an exponential rate $(\mathsf{C})$ and then with a power law rate $t^2$ $(\mathsf{D})$. 

\begin{figure}
\centering
\includegraphics[width=0.8\columnwidth]{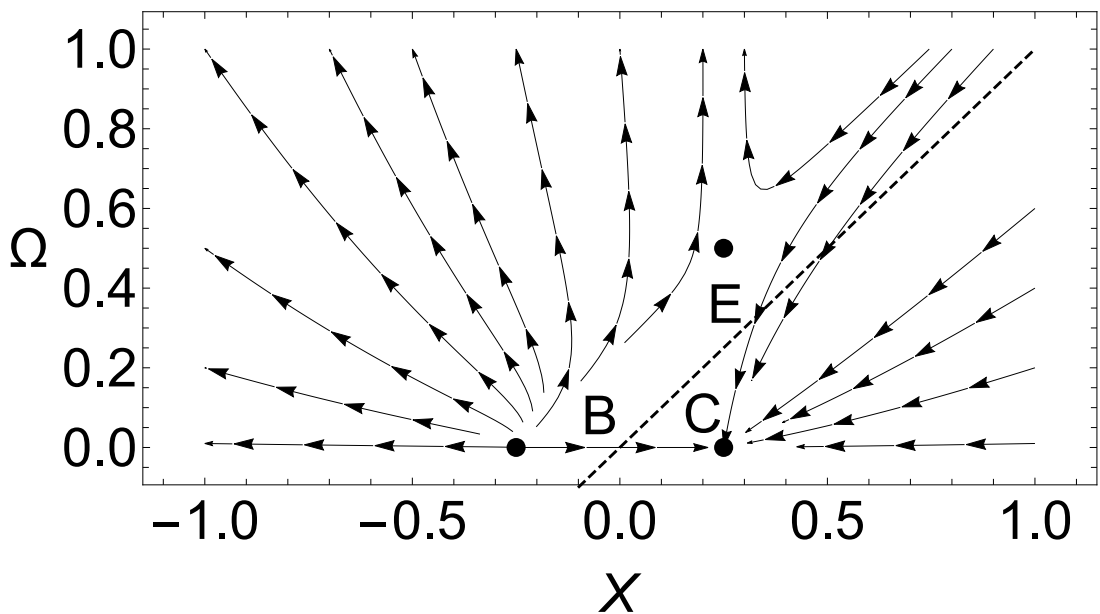}
\caption{The $K=0$, $Z=0$ invariant submanifold of the phase space for Equations A with the ansatz \eqref{Chi_ansatz_1}. Here $\mathbf A =1/3$ and the dashed line separates accelerating (right part of the plot) from decelerating expansion (left part of the plot).}
\label{Fig:EqA_2}
\end{figure}

The most interesting critical point of this system is Point~$\mathsf{D}$, the late time attractor where the universe undergoes an accelerated expansion provided that $(\mathbf{A}-1)(2\mathbf{A}-1) > 1$ which means $0<\mathbf{A}<1/2$. Once this choice has been made it is remarkable that `many' trajectories will make a transition from acceleration to deceleration and then back to acceleration before terminating at Point~$\mathsf{D}$. Considering also the invariant submanifold $K=0,Z=0$ (see Figure \ref{Fig:EqA_2}) we can explore   non vacuum orbits. From their behaviour we can conclude that the cosmology for action $S_1$ allows for early time and late time acceleration while at the same time allowing for a matter or radiation epoch.

\section{Conclusions}

We constructed a new class of modified theories of gravity which is characterised by a dynamical volume form. The key idea of this approach is to treat the volume form which appears in the action differently from the induced volume of the metric. This class of theories can be seen as a dynamical generalisation of `unimodular gravity' theories in which the volume form is a constant. The introduction of a variable volume form is achieved by the introduction of a fourth order tensor $\chi$ which connects the metric of the spacetime with another rank two tensor $\tilde{g}$ whose determinant expresses the dynamical volume form. No other modifications are introduced.

Our construction allows for two straightforward options in which matter couples to spacetime. First, we discussed the case where matter couples directly to the spacetime metric (theory $S_1$). Second, we considered the possibility where matter couples to $\tilde{g}$ (theory $S_2$). In this second case the gravitational field equations contain explicitly the matter Lagrangian, the cosmological dynamics implied by this model appear to be less interesting for cosmology.

Remarkably, in both cases a fairly general choice of $\chi$ leads to theories which contain only one additional scalar degree of freedom $\rho$. Theories of this type have been our main interest in the context of the evolution of cosmological spacetimes. In spite of their resemblance with standard scalar tensor gravity, our model presents some fundamental differences. For example, since $\rho$ is not a true independent dynamical variable, the theory does not present an independent equation for the evolution of this scalar degree of freedom. Note, however, the trace of the field equations amounts to a Klein-Gordon like equation  for $\rho$, much in the same way of one of the scalar field representation of $f(R)$ gravity.  

The phase space analysis of the theory $S_1$ presents some most interesting features which are almost unique when compared with other models. In particular, we can show that there exits a set of values of the parameters for which the universe is characterised by accelerated expansion ($q<0$) at early (inflation) and late times (dark energy). These two accelerating epochs are connected by an intermediate phase of decelerated expansion ($q>0$). Such features show that the new theory allows us to model the entire standard cosmology, from inflation to the dark era, including a phase of the matter or radiation domination. The phase space analysis also presents a Friedmannian fixed point $\mathsf{F}$, but the orbits associated to the cosmic histories of the accelerating type do not pass close to it. We expect, therefore, that in general our model will have matter eras which are different from $\Lambda$CDM cosmology. Such differences might generate signatures in some well known observables like the linear spectrum of structure. It is also remarkable that in general the double accelerating orbits will include {\it two} different accelerating eras: one unstable `almost' exponential expansion, and a stable power law one.    

In conclusion we found that a dynamical version of unimodular gravity with minimally coupled matter has, at least at the background level of cosmology, a number of interesting features which, in principle, have observable consequences. In this respect this class of model deserves further investigation not only at cosmological level, but also, for example, at astrophysical scales. Future works will be dedicated to such task.

\acknowledgments

SC was supported by an Investigador FCT Research contract through project IF/00250/2013 and acknowledges financial support provided under the European Union's H2020 ERC Consolidator grant ``Matter and strong-field gravity: New frontiers in Einstein's theory'' grant agreement No.~MaGRaTh646597, and under the H2020-MSCA-RISE-2015 grant No.~StronGrHEP-690904.

This article is partly based upon work from COST Action CA15117 (Cosmology and Astrophysics Network for Theoretical Advances and Training Actions), supported by COST (European Cooperation in Science and Technology).

\appendix
\section{The cosmology of action $S_2$}
\label{app:model2}
We include here, for completeness the analysis of the phase space of model $S_2$. Using the FLRW metric and the choice \eqref{Chi_ansatz_1}
 \begin{align}
   \left(2\rho ^2-1\right)\left(\dot{H}+H^2\right) &= -\frac{2\kappa  \mu }{3\rho }
   \left\{2 (\mathbf{A}-2) \rho ^2-w \left(2 \mathbf{A} \rho ^2-1\right) (\rho+3)+1 \right\}+
   \frac{H \dot{\rho} }{2 \rho }\left[(3\mathbf{A}-2) \rho ^2+1\right]+\mathbf{A}\ddot{\rho} \rho \,, \\
   \left(2\rho ^2-1\right)\left(H^2+\frac{k}{S ^{2}}\right) &=
   \frac{\kappa  \mu }{3 \rho } \left\{ -2 (\mathbf{A}-2) \rho ^2+
   \rho  w \left(2 (\mathbf{A}-1) \rho ^2+6 \mathbf{A} \rho +1\right)-2\right\}
   \nonumber \\ &+
   \frac{H \dot{\rho}}{2 \rho  }\left[2(3\mathbf{A}-2) \rho ^2-1\right] +
   \frac{ \ddot{\rho}}{2 \rho  }\left( 2 \mathbf{A}\rho ^2-1\right)\,, \\
   \dot{\mu}+3H (\mu+p) &= 0 \,,
 \end{align}
 and the final equation
\begin{multline}
  2\rho\left(2\rho ^2-1\right)\left(2 \mathbf{A}\rho ^2-1\right)\dddot{\rho} +
  \left[H B_1(\rho,\mathbf{A})+B_2(\rho,\mathbf{A})\dot{\rho}\right]\ddot{\rho} +
  H B_3(\rho,\mathbf{A})  \dot\rho^2  \\+
  \left\{2 \kappa  \mu  \left[B_4(\rho,\mathbf{A})+wB_{11}(\rho,\mathbf{A})\right] +
  \frac{2 k }{S^2} B_6(\rho,\mathbf{A})+ H^2 B_7(\rho,\mathbf{A}) \right\}\dot{\rho} \\ =
  2 H \kappa \mu  \left[B_8(\rho,\mathbf{A})-w B_{12}(\rho,\mathbf{A})+w^2 B_{13}(\rho,\mathbf{A}) \right] \,.
\end{multline}
Here
\begin{align}
  B_{11}(\rho,\mathbf{A}) &= -\frac{32}{3} \mathbf{A} [\mathbf{A} (6 \mathbf{A}-7)+2] \rho^7 -
  32 \mathbf{A} [\mathbf{A} (6 \mathbf{A}-7)+2] \rho ^6 +
  \left[\frac{40}{3} \mathbf{A} (2 \mathbf{A}-3)+16\right] \rho ^5 \\
  \nonumber &+
  [80 (\mathbf{A}-1) \mathbf{A}+32] \rho ^4+\frac{4}{3} (8 \mathbf{A}-9) \rho ^3+(12 \mathbf{A}-40)
  \rho ^2+\frac{4 \rho }{3}+10\,, \\
  B_{12}(\rho,\mathbf{A}) &= \frac{4}{3} \rho  \left(2 \rho^2-1\right)
  [2(\mathbf{A}-2) \rho^3 +12 (\mathbf{A}-1)\rho^2+3\rho+6]\,,  \\
  B_{13}(\rho,\mathbf{A})&= 2 (\rho +3) \left[8 \mathbf{A} \rho ^5-4 (\mathbf{A}+1) \rho ^3+2 \rho \right]\,.
   \end{align}
In the equations above we assumed that $L_{\rm (m)}=-p=-w\mu$ is the Lagrangian of a perfect fluid with barotropic equation of state $w$. 

\begin{figure}
\centering
\includegraphics[width=0.8\columnwidth]{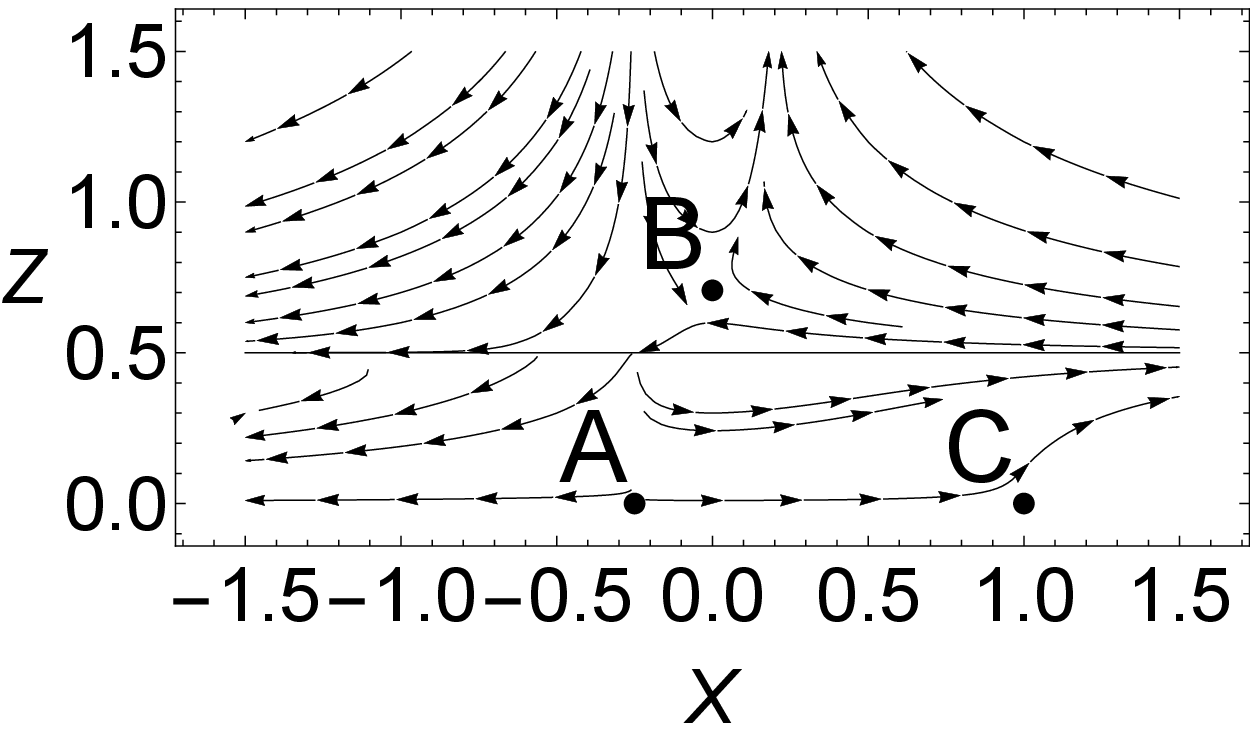}
\caption{The phase space for Equations B with the ansatz \eqref{Chi_ansatz_1} for spatially flat spacetime and vacuum. Here $\mathbf A =3$ and the dashed line separates accelerating (right part of the plot) from decelerating expansion (left part of the plot).}
\label{Fig:EqB_1}
\end{figure}

As before we can analyse the phase space of this cosmology using the variables
\begin{align}
X=\frac{\dot\rho}{4H\rho} \,,\qquad Y=\frac{\ddot\rho}{4H^2\rho}\,, \qquad Z=\rho \,,\qquad \Omega = \frac{\kappa \mu}{3 \rho H^2} \,,\qquad K=\frac{k}{ S^2 H^2} \,.
\end{align}
and the time variable $\tau=\ln (S/S_0)$. The dynamical equations  are then
\begin{align}\label{DynSys2}
\begin{split}
 X' &= \frac{1}{4 \mathbf{A} Z^2-2} \left\{K\left(Z^2 (2-4 \mathbf{A}
   X)-1\right)+4 X \left(2 \mathbf{A} w \Omega  Z^3+(2-3 \mathbf{A}) Z^2-w \Omega  Z+2 \Omega +1\right)\right.\\
   &~~~\left.+\Omega [ -4 \mathbf{A} w
    Z^3-(12 \mathbf{A} w -4 \mathbf{A}+8)Z^2 +2 w Z+6 w +2 
   ]-16 \mathbf{A} X^2 Z^2-2 Z^2+1\right\}\,,\\
Z' &= 4 X Z\,,\\
 K'&=\frac{4 K }{2 \mathbf{A}
   Z^2-1}\left[\mathbf{A} Z^2 (K-2 w \Omega  Z+1)+w \Omega  Z+2 X-2 \Omega -1\right]\,,\\
 \Omega' &=\frac{\Omega }{2 \mathbf{A} Z^2-1} \left[w \left(2 \mathbf{A} Z^2-1\right) (4 \Omega  Z-3)-4 X
   \left(2 \mathbf{A} Z^2+1\right)-4 \mathbf{A} K Z^2-6 \mathbf{A} Z^2+8 \Omega +5\right]\,.
   \end{split}
\end{align}
where the prime denotes the derivative with respect to $\tau$ and we applied the constraint
\begin{multline}
  \Omega  \left[2 (\mathbf{A}-2) Z^2-w (Z+3) \left(2 \mathbf{A} Z^2-1\right)+1\right]+X \left[(4-6 \mathbf{A}) Z^2+1\right] \\ +
  Y \left(1-2 \mathbf{A} Z^2\right)+\frac{1}{2} (K-1) \left(2 Z^2-1\right)=0 \,.
\end{multline}
The solutions associated to the fixed points can be found solving the equation
\begin{align}
  H'= \frac{H}{1-2 \mathbf{A} Z_*^2}\left\{K_*-1+\Omega_*  \left[4-2 w
    Z_*(2 \mathbf{A} Z_*^2-1)\right]+4 X_*+K_*(2 \mathbf{A} Z_*^2-1)\right\} \,,
\end{align}
where an asterisk denotes the value of the variables in the fixed point. As before, the physical phase space will be characterised by $Z>0$ and the system presents the invariant submanifolds $\Omega=0$, $K=0$ and $Z=0$ so that a global attractor will necessarily have coordinates $\Omega=0, K=0, Z=0$. The system \eqref{DynSys2} also present the same singular manifold as the previous set of equations in $Z=(2A)^{-1/2}$. In our treatment we will consider only $Z\neq(2A)^{-1/2}$. These analogies should not be surprising, as the two sets of field equations only differ in the matter sector. The fixed points and their stability can be found in Table \ref{Tab:EqB_1}. Differently  from the previous case we now have only isolated fixed points of which only two ($\mathsf{B}$ and $\mathsf{D}$) can be attractors, neither of them global. Point~$\mathsf{B}$ can represent accelerated expansion if $\mathbf{A}>2$. In Fig.~\ref{Fig:EqB_1} we give a plot of the phase space for $K=0$, $\Omega=0$ for $\mathbf{A}=3$.

\begin{table}
\begin{center}
\begin{tabular}{ccccc}
 Point & Coordinates & Attractor & Repeller & Solutions \\\hline 
$\mathsf{A}$  & $\left\{K\to 0,\Omega \to 0,X\to -\frac{1}{4},Z\to 0\right\}$ & Always & Never & $a\to a_0 \sqrt{ t-t_0}$ \\ \\
\multirow{2}{*}{$\mathsf{B}$} & \multirow{2}{*}{$\left\{K\to 0,\Omega \to 0,X\to 0,Z\to \frac{1}{\sqrt{2}}\right\}$} & $\mathbf{A}>2$ & $1< \mathbf{A}<\frac{4}{3}$ & \multirow{2}{*}{$a\to a_0 \left(t-t_0\right)^{\mathbf{A}-1}$} \\ 
& & $\mathbf{A}\neq \mathbf{A}_{0}^{+}$ & $\mathbf{A}\neq \mathbf{A}_{0}^{-}$\\ \\
$\mathsf{C}$  & $\{K\to 3,\Omega \to 0,X\to \frac{1}{2},Z\to 0\}$ & Never & Never & $a\to a_0 \left(t-t_0\right)$ \\ \\
$\mathsf{D}$  & $\left\{K\to \frac{2-\mathbf{A}}{\mathbf{A}},\Omega \to 0,X\to 0,Z\to\frac{1}{2}\right\}$ & $1<\mathbf{A}<2$ & Never & $a\to a_0 \sqrt{ t-t_0}$\\ \\
$\mathsf{F}_i$  & $\left\{K\to 0,\Omega \to \frac{6 \mathbf{A} w Z_{0,i}^2+6 \mathbf{A} Z_{0,i}^2-3 w-5}{4
   \left(2 \mathbf{A} w Z_{0,i}^3-w Z_{0,i}+2\right)},X\to 0,Z\to Z_{0,i}\right\}$ & Never & Never & $a\to a_0 \left(t-t_0\right)^{\frac{2}{3(1+w)}}$  \\ \\ \hline \\
\multicolumn{5}{c}{$\frac{1}{w Z_{0,i} \left(2 \mathbf{A} Z_{0,i}^2-1\right)+2}\left\{2 w Z_{0,i}^3 [3 \mathbf{A} (w+1)+2]+6 (w+1) Z_{0,i}^2 [\mathbf{A} (3 w-1)+2]-w (3 w+7) Z_{0,i}-9 w(w+2)-1\right\}=0$}\\ 
\multicolumn{5}{c}{$\mathbf{A}_{0}^{\pm}=\frac{1}{9} \left(20\pm4\sqrt{7}\right)$}\\
   \\\hline
\end{tabular}
\end{center}
\caption{Critical points stability and associated solution of Equations B with the ansatz \eqref{Chi_ansatz_1}. The index $i$ of the points $\mathsf{F}_i$ runs from 1 to 3 and $Z_{0,i}$ are the real solutions of equation in the last row. }
\label{Tab:EqB_1}
\end{table}

\clearpage

\providecommand{\href}[2]{#2}\begingroup\raggedright\endgroup

\end{document}